# ON PERMANENT AND SPORADIC OSCILLATIONS OF THE MAGNETOSPHERE

A.V. Guglielmi[1] and A.S. Potapov[2*]

[1]*Institute of Physics of the Earth of RAS, Moscow, Russia*

[2]*Institute of Solar-Terrestrial Physics of SB RAS, Irkutsk, Russia*

*Corresponding author:* **A.S. Potapov,** Institute of Solar-Terrestrial Physics of Siberian Branch of Russian Academy of Sciences, P.O.Box 291, Irkutsk, 664033 Russia, e-mail: potapov@iszf.irk.ru; tel.: +7 3952 511673

**Abstract.** In this paper we investigate the impact of permanent oscillations Pc3 on the excitation of sporadic oscillations Pi2 (their periods are 10–45 and 40–150 s, respectively). The hypothesis is formulated that Pc3 oscillations originating in front of the magnetosphere penetrate into the geomagnetic tail, cause a local depression in the current in the neutral sheet, and under favorable conditions stimulate a tearing instability. This leads to reconnection of magnetic field lines and an explosive release of magnetic energy stored in the tail. As a result, a substorm breaks up, with sporadic pulsations Pi2 as an important element of this process. It is expected from theoretical estimates and kinematic considerations that the higher the Pc3 frequency, the earlier the Pi2 trains start. We test this prediction using observational data from satellite measurements of the interplanetary magnetic field and on-ground magnetic measurements. The results confirm the theoretical expectation. Additional routes are proposed to theoretically and experimentally test the hypothesis.

**Introduction**

In geophysics, a morphological classification of ultra-low frequency (ULF) oscillations of the magnetosphere has been accepted since the mid-sixties of the last century (see review [Troitskaya, Guglielmi, 1967]). Various oscillation modes are subdivided into two classes: Pc (continuous pulsations) and Pi (irregular pulsations). This paper discusses the basic subclasses of Pc and Pi, namely, the daily permanent Pc3 oscillations and sporadic night Pi2 oscillations (period ranges are 10–45 and 40–150 s, respectively).

Pc3s and Pi2s appear not to be cause-and-effect related, as their respective origins lie in the near-Earth space areas separated by 150 thousand kilometers and more. A comparative analysis of the oscillations presented below, however, reveals a possible impact of Pc3 oscillations on the excitation conditions of Pi2. More specifically, Pc3 fluctuations may affect the reconnection process of geomagnetic field lines, leading to Pi2 excitation.

The idea of reconnection was proposed by R.G. Giovanelli in relation to the problem of the solar flare interpretation [Giovanelli, 1946]. J.W. Dungey pointed to the fundamental role of the geomagnetic field line reconnection in the process of magnetic storms [Dungey, 1958]. The reconnection problem is the subject of extensive literature on geophysics, astrophysics and plasma physics [Kadomtsev, 1987; Zeleny, Milovanov, 2004; Zeleny et al., 2010; Yamada et al., 2010]. However, the key questions of the theory remain unsolved. They continue to cause considerable debate (see for example [Goldstein, 2001]). Moreover, the problem of reconnection is listed among the main outstanding problems of modern physics (see http://List_of_unsolved_problems_in_physics). On the other hand, the phenomenology of reconnection is rather simple and physically transparent, and we use this for a qualitative study of the relationship between the Pc3 and Pi2 oscillations.

First, we remind the known information on the origin of these two types of geomagnetic pulsations. Then we shall state our hypothesis about Pc3 as a trigger for Pi2 excitation in the geomagnetic tail. After that we shall use observational data to test the conclusions derived from the theory. Finally, we shall discuss the results and outline some directions for future work.

**Origin of Pc3 and Pi2 oscillations**

The permanent oscillations Pc3 exist as quasi-periodic geomagnetic field pulsations. They originate in front of the magnetosphere due to the ion-cyclotron instability of the solar wind protons, which are partially reflected from the front of the Earth's bow shock [Guglielmi, 1974]. Sporadic Pi2 oscillations arise as a short train resulting from an explosive process in the magnetotail [Guglielmi,

Troitskaya, 1973; Nishida, 1978]. What both the permanent and sporadic oscillations have in common is that their properties essentially depend on the interplanetary magnetic field (IMF). In the language of the theory of dynamical systems and catastrophe theory [Gilmore, 1984], the IMF components constitute a set of driving parameters, the combination of which controls the oscillation mode.

The constantly blowing solar wind steadily supports the excitation of Pc3. Originating ahead of the near-Earth bow shock front the Pc3 waves penetrate into the magnetosphere and are observed almost continuously on the sunward side of the globe. The oscillation frequency is controlled by the absolute value of interplanetary magnetic field |**B**|. The amplitude of the oscillations depends, in a specific manner, on the angle $\phi$ between the IMF field line and vector of the solar wind velocity. On average, $\phi \approx \pi/4$ at the Earth's orbit, but sometimes $\phi$ deviates considerably from this value.

Here, it should be specified that, in the standard model, the quasi-monochromatic Pc3 oscillations are a result of selective amplification of broadband magnetosonic fluctuations inherent in the solar wind [Guglielmi, 1974]. The amplification occurs in the foreshock or, to be more precise, in the area extending from the bow shock front to the front edge of the reflected protons. The latter position is controlled by angle $\phi$. Based on geometrical considerations we tentatively have $R_{\parallel} \approx \text{const} \cdot R_{\perp} \cot \phi$ where $R_{\parallel}$ is the maximum length of wave amplification, $R_{\perp}$ is the effective radius of the magnetosphere cross-section, and the constant is on the order of unity. This formula is applicable when $R_{\parallel} \ll L$. Here $L$ is the mean free path of the reflected protons. This length is determined by proton scattering by magnetosonic fluctuations. Thus, when $\phi \to 0$ the formula does not work. When $\phi \to \pi/2$, however, it is applicable, and specifically indicates the disappearance of Pc3 at the moment when IMF vector **B** becomes orthogonal to the solar wind flow direction. With this IMF configuration, Pc3 oscillations suddenly stop.

There was a time when the sudden disappearance of permanent oscillations at all observatories of the daytime hemisphere caused considerable astonishment, and the natural intention to produce a more or less arbitrary interpretation for poorly understood phenomena has given rise to a number of barely plausible theories (see for example [Troitskaya, Guglielmi, 1969]). After the above-outlined Pc3 theory was created, however, the sudden cessation of oscillations when $\phi \to \pi/2$ became entirely obvious.

Thus, |**B**| and $\phi$ are the parameters determining the mode of the permanent oscillations. An additional control parameter is the solar wind velocity [Guglielmi, Potapov, 1994]. It affects the

intensity of the oscillations. In the overall picture, there are no signs of any catastrophic behavior of the oscillatory system, apart from rare cases of sudden Pc3 disappearance.

The situation is different for sporadic fluctuations. For Pi2, the control parameter is the component of **B** that is orthogonal to the ecliptic plane. It is the $B_z$ component in the Geocentric Solar Magnetospheric (GSM) coordinate system. Usually, the Pi2 oscillations arise shortly after $B_z$ switches from positive to negative values. Recall that $B_z > 0$ ($B_z < 0$), if the **B** vector is directed towards the northern (southern) hemisphere.

After $B_z$ reversal a large-scale catastrophe occurs in the magnetosphere. The IMF field lines are interconnecting with the geomagnetic field lines in the noon sector of the magnetopause when the $B_z$ sign changes from positive to negative. The reconnected lines are transported by solar wind into the magnetospheric tail. Magnetic energy in the tail increases and, at some point, opposite-directed field lines are rapidly reconnecting through the neutral sheet of the tail. The magnetic energy stored in the tail is explosively converted into the kinetic energy of charged particles. The magnetosphere deforms and starts to vibrate; energetic electrons are invading the upper atmosphere, causing a bright glow; the current along the auroral zone in the ionosphere increases sharply; the current fluctuations excite atmospheric infrasonic waves, etc. Together, these phenomena are called a magnetospheric substorm [Nishida, 1978]. Sporadic Pi2 oscillations are an important element of the substorm. From the perspective of a terrestrial observer the substorm usually begins with an appearance of a powerful short Pi2 train that is observed in the night hemisphere.

A small digression is appropriate here so as to point out the relevance of improving the classification of ULF oscillations. Proposed 50 years ago, the morphological classification should be supplemented with elements of the genetic classification based on certain conceptions regarding the oscillation excitation mechanisms. This will help to translate a significant part of empirical material gained during the many years of ULF research into the universal language of critical phenomena, phase transitions and catastrophe theory. A physics-based systematization of oscillations will enhance understanding between researchers investigating critical phenomena in the magnetosphere. Besides, it will allow finding the useful analogies and facilitate the creation of generalized models of geophysical catastrophes, which is especially important in solving problems involving interactions between the geospheres.

**Pc3 oscillations as a trigger for magnetic reconnection**

Thus, the key role in both Pi2 initiation and substorm development is played by magnetic field line reconnection. The reconnection first occurs in the dayside magnetopause, and somewhat later in the neutral layer of the magnetotail. Topologically, the reconnection process is clear, and the above described sequence of events seems plausible. But if so, what is the subject of the problem of reconnection? The problem attracts widespread interest, but cannot be solved for nearly 70 years.

The matter is as follows: On the one hand, the reconnection should proceed rapidly. This is evidenced by astrophysical and geophysical observations as well as results of laboratory experiments and computer simulations. On the other hand, the theory implies an excessively slow rate of reconnection. Reconnection rate is characterized by dimensionless Mach-Alfven number $M_A$. In the Sweet-Parker theory, $M_A \sim R_m^{-1/2}$, where $R_m$ is the magnetic Reynolds number [Parker, 1957]. But $R_m \gg 1$ in all known cases of reconnection in space plasma. The value of $M_A$ appears to be unacceptably low. In the well-known Petscheck model, the situation is somewhat better: $M_A \sim (\ln R_m)^{-1}$ [Petschek, 1964]. However, the quantitative discrepancy remains. In addition, some conceptual difficulties have been gradually revealed in the Petscheck theory over the years [Kadomtsev, 1987; Goldstein, 2001; Yamada et al., 2010].

Our task does not involve providing a detailed discussion or even solving the reconnection problem. We restrict ourselves to outlining the idea of Pc3 as a probable trigger for field line reconnection through the neutral layer of the magnetotail.

After crossing the magnetopause, Pc3 waves do not only fall on Earth, but also penetrate into the geomagnetic tail. When propagating along the tail as they would along a tube, Pc3 waves tend to concentrate in the vicinity of the neutral layer dividing the tail into the northern and southern halves. Recall that in the northern (southern) half field **B** is in the sunward (antisunward) direction, and that a strong current flows along the neutral sheet from the dawn to the dusk side. (To avoid confusion, it must be pointed out that, in contrast to the previous section, **B** is geomagnetic field in the tail in this one.) With increasing distance from the Earth, the field lines of the northern and southern halves of the tail converge at an acute angle when approaching the neutral layer. Such a field configuration provides for the wave energy to be concentrated in the vicinity of the neutral sheet if Pc3 propagate as the Alfven waves. We restrict ourselves to analyzing this particular case. Note, however, that the fast magnetosonic waves will also propagate toward the neutral layer due to refraction at the boundary of the plasma layer into which the neutral layer is immersed.

We need to specify the model of magnetic field **B**. The simplest model has the form $B_x(z) = -B_0 \text{sign}(z)$, $B_y = B_z = 0$. Here and below, the $x$ axis is antisunward along the tail, $y$ is along the neutral plane $z = 0$, and the z axis is perpendicular to the neutral plane and directed northward. From the equation $\mathbf{j} = (c/4\pi) \text{rot} \mathbf{B}$ we find the nonvanishing component of the current flowing duskward along the neutral plane: $j_y(z) = -(c/2\pi)B_0\delta(z)$. This model is useful for studying the tearing instability of plane-parallel currents, presumably leading to rapid reconnection. For our purposes, however, the model should be supplemented, to more closely reflect the reality.

First, let us introduce finite thickness of the neutral layer. For this purpose we use the well-known Harris model (see reviews [Kadomtsev, 1987; Zeleny et al, 2010]). Second, based on satellite observations, let us add a nonzero field component that is orthogonal to the neutral layer. As a result, the field components will acquire the following form: $B_x(z) = -B_0 \tanh(z/\Delta)$, $B_y = 0$, $B_z = B_n$. When $|z| \gg \Delta$, the field lines are almost straight in each half of the tail and converge toward the neutral layer at an acute angle $\sim 2B_n/B_0$, $B_n \ll B_0$. When $|z| \ll \Delta$, the field lines are parabolic in shape:

$$x = x_0 - \frac{B_0 z^2}{2 B_n \Delta}. \qquad (1)$$

The radius of field line curvature is minimum in the $z = 0$ plane and is equal to $(B_n/B_0)\Delta$, i.e. much smaller than the layer thickness. Thus, the Alfven wave propagating along the tail reaches the neutral layer.

Let us discuss how the wave interacts with the neutral layer. For the purpose, let us use the canonical form of the one-dimensional wave equation [Ginzburg, 1960]. In Maxwell's equations, let us set $\partial/\partial t = -i\omega$, $\partial/\partial x = \partial/\partial y = 0$, obtaining the equation of a plane monochromatic wave normally incident on the neutral layer:

$$E_x'' + k^2(z) E_x = 0. \qquad (2)$$

Here the prime denotes differentiation with respect to $z$, $k = (\omega/c)n$ is the wave number, $\omega$ is wave frequency, $n$ is the refractive index, $c$ is the velocity of light. The square of the refractive index of the Alfven wave propagating at a large angle to magnetic field lines is

$$n^2 = \left(\frac{B}{B_n}\right)^2 \left(1 - \sum \frac{\omega_0^2}{\omega^2 - \Omega^2}\right), \qquad (3)$$

where $B = \sqrt{B_x^2 + B_n^2}$ is the absolute value of the magnetic field vector, $\omega_0 = (4\pi e^2 N/m)^{1/2}$ is plasma frequency, $\Omega = eB/mc$ is cyclotron frequency; $e$, $m$, and $N$ are the charge, mass and number density of the species of particles (see [Guglielmi, Potapov, 2012]). The summation in (3) is over all species of particles (electrons and ions). The wave is polarized as follows: $\mathbf{E}=(E_x, 0, E_z)$, $\mathbf{b}=(0, b_y, 0)$. Thus, $E_z = -(B_x / B_n) E_x$, $b_y = (ic/\omega) E_x'$. Here we took into account inequality $\omega n \ll \omega_{0e}$ which is easily satisfied for Pc3 waves (index $e$ indicates electrons). The relation between the components of the Poynting vector $\mathbf{S} = (c/4\pi) \mathbf{E} \times \mathbf{b}$ follows directly from the above formulas: $S_x / S_z = B_x / B_n$, $S_y = 0$, and $|S_x| \gg |S_z|$ everywhere, except for a layer of thickness $\delta = (B_n / B_0) \Delta$.

It is also necessary to set the unperturbed plasma distribution. Plasma in the tail contains protons and alpha particles of solar origin, as well as an admixture of oxygen ions of ionospheric origin. A multicomponent plasma composition and specific configuration of magnetic field indicate that, in the neutral layer, there is a so-called ion-cyclotron resonator similar to a cavity in the equatorial region of the radiation belt [Guglielmi et al., 2000; Guglielmi, Potapov, 2012]. Generally speaking, we cannot exclude the possibility that the resonant excitation of ion-cyclotron waves in the neutral layer could be linked to the reconnection problem. An appropriate analysis would take us too far afield, however. For simplicity's sake, let us consider here a pure hydrogen plasma. In this case, an ion cyclotron resonator does not occur. Instead of cumbersome formula (3) we have

$$n = \left(\frac{B}{B_n}\right) \frac{\omega_{0p}}{\sqrt{\Omega_p^2 - \omega^2}}, \qquad (4)$$

where the $p$ index means that the parameter refers to protons. Dependence $n(z)$ is determined by the two functions $B_x(z) = -B_0 \tanh(z/\Delta)$ and $N(z) = N_0 \operatorname{sech}^2(z/\Delta) + N_\infty$. Here we used the formula for $N(z)$ from [Treumann, Baumjohann, 2013], adding term $N_\infty \ll N_0$ to the right-hand side of equation.

We are interested in the force and heat impact the wave exerts on the neutral layer. Strictly speaking, this requires a nonlinear self-consistent problem to be solved. In this paper we restrict ourselves to a preliminary analysis of the impact of the ponderomotive force $\mathbf{F}$ associated with the wave on the current $\mathbf{j}$ in the neutral layer. We rely on the results of a study of ponderomotive forces in space plasmas detailed in the review by [Lundin, Guglielmi, 2006].

Let us focus on the question of which part of the tail should expect the maximum ponderomotive action of Pc3 on the neutral layer current. The general expression for $\mathbf{F}$ contains a number of components, namely, the Abraham force, the Miller force, the Pitaevski force, and so on. A comparative analysis shows that in the above model, the Pitaevski force will dominate in that part of the tail where $\omega \sim \Omega_p$. This force is proportional to the square of the wave amplitude and directed to the $z=0$ plane, in both the northern and the southern half of the tail. In other words, force **F** is nearly perpendicular to field **B** throughout, except for a layer of small thickness $\delta \ll \Delta$.

The cosmic electrodynamics equations [Alfven, Felthammar, 1967] show that in crossed **F** and **B** fields the charged particles drift at velocity

$$\mathbf{V} = \frac{c}{eB^2} \mathbf{F} \times \mathbf{B}. \tag{5}$$

We can see that electrons ($e<0$) and protons ($e>0$) drift in opposite directions, creating a current $\delta \mathbf{j}$. Given the **F** properties described above, and the **B** model we have adopted, we find that $\mathbf{j} \cdot \delta \mathbf{j} < 0$. In other words, the ponderomotive force-induced current is directed from the dusk to the dawn part of the tail and thus the total current in the neutral layer decreases.

It is easy to see that the above decrease in the current occurs at a well-defined distance $x_*(\omega)$ from the Earth, and the lower the Pc3 oscillation frequency, the larger $x_*$. Indeed, the value of $B = \sqrt{B_n^2 + B_x^2}$ slowly decreases with increasing $x$. The cyclotron frequency $\Omega_p(x)$ decreases in exactly the same manner. At low $x$, i.e. at relatively small distances from the Earth, we have $\omega \ll \Omega_p$. Here the non-resonant Miller force dominates. Ponderomotive resonance $\Omega_p(x) \to \omega$ occurs at a well-defined distance $x \to x_*(\omega)$ that increases with decreasing oscillation frequency $\omega$. If we put $B \propto x^{-\mu}$, where $\mu$ is on the order of unity, then $x_* \propto \omega^{-1/\mu}$. Thus, the picture is quite clear qualitatively. As for quantitative calculations, they would require a more detailed numerical study. For reference, we give here the $B_n(x) \propto x^{-0.8}$ and $B_0(x) \propto x^{-0.3}$ dependences, which can be found in the literature describing magnetic field structure of the tail. However, it should be kept in mind that the power exponents of distance change with time in accordance with the changing conditions of the solar wind flowing around the magnetosphere.

Thus, we have set out everything that is necessary to make the following hypothesis. Permanent Pc3 oscillations cause local reduction of the neutral layer current at $x \sim x_*(\omega)$, which under favorable conditions, stimulates a tearing instability and leads to rapid reconnection of field lines in the magnetotail. For the hypothesis to be justified theoretically, the problem of self-consistent nonlinear interaction of Alfven waves with a neutral layer must be solved. In a sense, it is more interesting firstly to verify the above inferences experimentally.

**Experimental test**

To test our hypothesis concerning the role of geomagnetic Pc3 pulsations in stimulating magnetic reconnection in the magnetotail, we used satellite measurements of the interplanetary magnetic field and the solar wind velocity, using the one minute resolution database OMNI (http://omniweb.gsfc.nasa.gov/form/omni_min.html) and ground-based observations of geomagnetic pulsations at mid-latitude observatory Uzur (53.3°N, 107.4°E) for the period 1 November 2009 to 15 April 2010. We used magnetic field oscillation records obtained from the $H_x$ channel output of a LEMI-418 induction magnetometer. During this period 79 Pi2 trains with amplitudes exceeding 40 pT were registered, observable from 3 pm to 5 am local time (0800 to 2200 UT). From these, we excluded the 8 trains at least five hours before which no southward reversal was observed in the IMF vertical component. There are no IMF or solar wind parameter measurement data for 5 events. Three Pi2 trains were observed on the background of high speed solar wind streams under strong magnetic disturbances – these were also excluded. Finally, we discarded another 11 events in which $B_z$ did not drop below –1.6 nT between the time the IMF $B_z$ turned southward and Pi2 onset. The remaining 52 Pi2 trains were tabulated into Table 1, where the oscillation onset date and time (UT) are indicated, as well as the following information: maximum $B_z$ one hour before this component turned southward; minimum $B_z$ during the interval between the time this component turned southward and the Pi2 train started; duration of this interval ($\Delta t$), in minutes; mean absolute value of the IMF ($|\mathbf{B}|$) and the solar wind velocity ($V_{sw}$) within the specified interval.

Next, we used the following scheme of analysis. The typical sequence of events before the substorm is known to include:

- a southward reversal of the IMF vertical component $B_z$;

- $B_z$ remaining negative for some time $\Delta t$ (fractions of an hour to several hours), during which field lines from the dayside magnetosphere move to the magnetotail;

- magnetic reconnection in the magnetotail leading to Pi2 pulsations generated in the train, which coincides with the substorm onset.

Using data from satellite and ground-based observations in Table 1, we tried to find a connection between the Pc3 carrier frequency and the time lag between the $B_z$ southward turn and the Pi2 onset. Since Pi2 are a predominantly nighttime phenomenon, however, and Pc3 are dawn-daytime oscillations, they cannot be observed simultaneously at one and the same observatory. Therefore, we evaluated the Pc3 carrier frequency from the IMF magnitude $|\mathbf{B}|$. The relationship between these two parameters is well known [Guglielmi, 1974]: $f \approx g|\mathbf{B}|$, where constant $g$ has repeatedly been determined from the extensive statistical material. Since we do not search for a precise expression of the relationship between the Pc3 frequency and the time lag of substorm onset, but only try to detect the very existence of such a relationship, we take the following – the most frequently used – value of the constant: $g = 6$ mHz/nT. In the last column of Table 1, we estimate the carrier frequency of Pc3 for all events. We compared the expected Pc3 frequency with the measured delay of the Pi2 start relative to the southward $B_z$ turn. The result in Fig. 1 suggests a clear tendency in the delay time to increase with decreasing Pc3 frequency. The correlation coefficient is –0.58 for the linear regression (dashed line), and –0.66 for the power regression (dash-dotted line). The power regression equation has the form $\Delta t \approx 8 \cdot 10^2 f^{-0.67}$, where the delay time $\Delta t$ is measured in minutes, and the Pc3 frequency in millihertz.

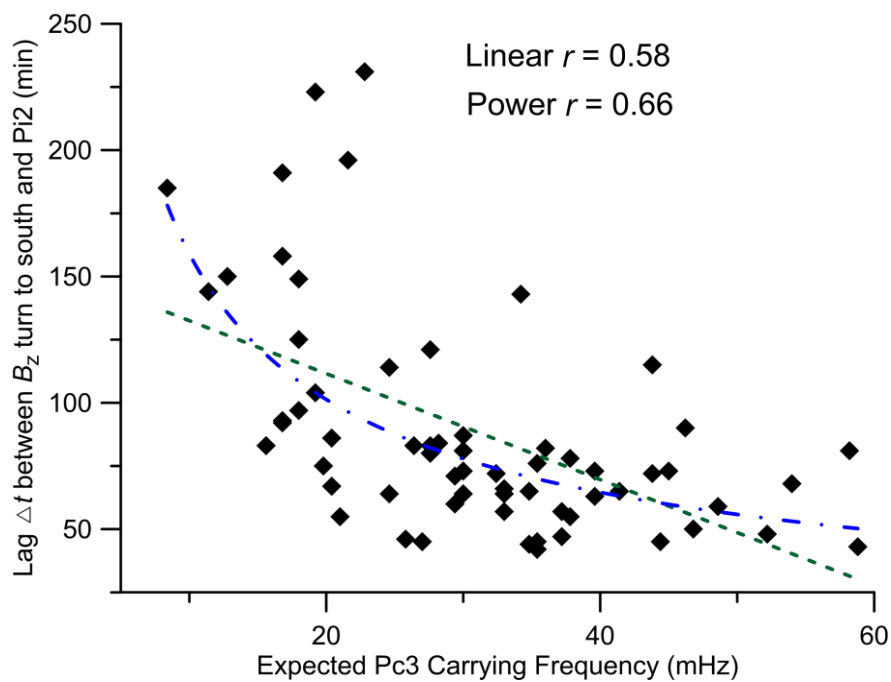

**Figure 1.** Dependence of a time delay $\Delta t$ between Pi2 train onset and $B_z$ southward reversal on the expected Pc3–4 frequency.

The question may be raised, however, of whether the observed effect is merely a consequence of the delay depending on the $B_z$-component, which, in turn, is connected to the IMF magnitude $|\mathbf{B}|$. In this case, Pc3 oscillations would be redundant in this causal chain. The plot in Fig. 2$a$ allows us to reject this assumption. Indeed, it shows an almost complete lack of any dependence of time delay $\Delta t$ on the IMF vertical component, despite the fact that some connection of the IMF magnitude with the $B_z$-component does exist, as seen in Fig. 2$b$.

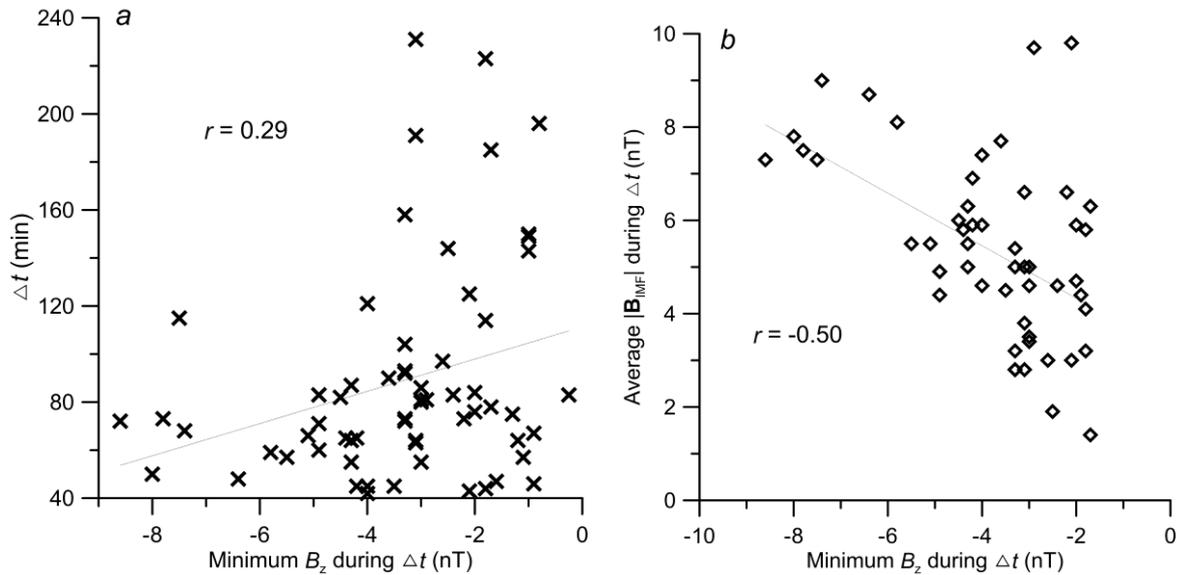

**Figure 2.** (*a*) Pi2 onset delay $\Delta t$ dependence on the minimum $B_z$-component within $\Delta t$; (*b*) the average IMF magnitude during the $\Delta t$ interval versus the minimum $B_z$-component within $\Delta t$.

**Discussion and Conclusion**

The main objective of the study is to stimulate discussion on a role of the permanent geomagnetic pulsations in the reconnection of field lines in the magnetotail. We put forward the hypothesis that the permanent Pc3 oscillations exert a stimulating influence on the excitation of sporadic Pi2 fluctuations.

Our theoretical analysis shows that Pc3 waves produce in the magnetotail a ponderomotive force that causes a local reduction in the neutral layer current. Under favorable conditions, this stimulates a tearing instability that leads to rapid field line reconnection in the magnetotail. A substorm breaks out as a result; its onset signaled by a train of sporadic Pi2 oscillations. The Pc3 impact is strongest at the point of so-called ponderomotive resonance, where the Pc3 frequency approaches proton gyrofrequency. Since magnetic field decreases with distance from the Earth, the point of the ponderomotive resonance moves farther away with decreasing Pc3 frequency.

In terms of kinematics, the above model suggests that the higher the Pc3 frequency, the shorter the delay between the reorientation of the IMF vertical component at the magnetospheric front and the time of Pi2 onset at observatories in the night-time hemisphere. We test this conclusion by comparing satellite data with Pi2 observations at one mid-latitude observatory. The results show a clear tendency for an inverse relationship between the above delay and the IMF magnitude that is proportional to the Pc3 frequency.

The original problem could be formulated differently. Namely, it is natural to ask the question: what determines the time delay between the $B_z$ inversion and observed Pi2? In these terms, we would first check if the delay depends on the minimum value of the $B_z$ component after its inversion but before the Pi2 onset, because, at first glance, a large negative $B_z$ means a high reconnection rate at the dayside magnetopause and a large magnetic flux thrust into the magnetotail. In turn, large magnetic flux in the tail must lead to earlier reconnection through the neutral layer. But experiment show that $\Delta t$ does not depend on minimum $B_z$. Then we would look for other possible sources affecting $\Delta t$ and find that $\Delta t$ depends on $|\mathbf{B}|$. But why? Apparently, the large magnetic flux transferred from the dayside is not sufficient for the reconnection to begin in the tail, some additional conditions are needed. Among these possible conditions, wave activity is very important because reconnection requires high anomalous resistance, which in collisionless plasma can be provided by wave turbulence [Priest, Forbes, 2007]. Pc3 waves penetrating into the geomagnetic tail can play the role of such turbulence. Their frequency is proportional to the IMF magnitude, so the use of the above theory helps us to arrive at an explanation of the $\Delta t$ dependence on $|\mathbf{B}|$.

*Acknowledgements.* The OMNI data were obtained from the GSFC/SPDF OMNIWeb interface at http://omniweb.gsfc.nasa.gov. The authors acknowledge J. King, N. Papitashvili and R. McGuire as persons responsible for data providing and operation of OMNI website. This investigation was performed in the framework of Project 6.2 of Program 4 and Project 22-4 of Program 22 of RAS Presidium and RFBR projects 13-05-00066 and 13-05-00529.